\documentclass[twocolumn,aps,superscriptaddress,nofootinbib,floatfix,linenumbers]{revtex4}
\usepackage{ulem}
\usepackage{color}
\usepackage{soul,color}
\usepackage{graphicx}
\usepackage{bm}
\usepackage{amsfonts}
\usepackage{float}
\usepackage{placeins}
\usepackage[hidelinks]{hyperref}
\usepackage{color,amsmath,amssymb,bm}
\usepackage{tensor}
\usepackage[cmyk]{xcolor}
\newcommand{\be}{\begin{equation}}
\newcommand{\ee}{\end{equation}}
\newcommand{\ba}{\begin{eqnarray}}
\newcommand{\ea}{\end{eqnarray}}

\begin{document}
\title{Bottomed mesons and baryons in pp collisions at $\sqrt{s}=5 \, TeV$ LHC energy within a Coalescence plus Fragmentation approach}
\author{Vincenzo Minissale}
\email{vincenzo.minissale@dfa.unict.it}
\affiliation{Dipartimento di Fisica e Astronomia "E. Majorana", Università degli Studi di Catania, Via S. Sofia 64, 1-95125 Catania, Italy}
\affiliation{INFN Sezione di Catania, Via Santa Sofia,64 - 95123 Catania, Italy}

\author{Vincenzo Greco}
\email{greco@lns.infn.it}
\affiliation{Dipartimento di Fisica e Astronomia "E. Majorana", Università degli Studi di Catania, Via S. Sofia 64, 1-95125 Catania, Italy}
\affiliation{Laboratori Nazionali del Sud, INFN-LNS, Via S. Sofia 62, I-95123 Catania, Italy}

\author{Salvatore Plumari}
\email{salvatore.plumari@dfa.unict.it}
\affiliation{Dipartimento di Fisica e Astronomia "E. Majorana", Università degli Studi di Catania, Via S. Sofia 64, 1-95125 Catania, Italy}
\affiliation{Laboratori Nazionali del Sud, INFN-LNS, Via S. Sofia 62, I-95123 Catania, Italy}
\begin{abstract}
Recent experimental data from $pp$ collisions have shown a significant increase in heavy baryon production leading to a baryon over meson ratio which is one order of magnitude higher than elementary collisions ($e^+e^-$, $ep$). From a theoretical point of view this large production of baryon can be explained with hadronization via quark coalescence assuming a QGP medium in $pp$ collisions.
In this study, we extend this analysis to include hadrons containing bottom quarks. Employing a coalescence plus fragmentation approach, we present predictions for $p_T$ spectra and the heavy baryon/meson ratio of charmed hadrons with and without strangeness content, specifically: $\bar{B^0}$, $B_s$, $\Lambda_b$, $\Xi_b^{0,-}$, $\Omega_b$, and the $B_c$ meson. 
We have found that coalescence is the dominant mechanism in the B meson production, especially at low momenta, at variance with what found in the charm sector where the D meson were mainly produced via fragmentation.
Our model predicts a $\Lambda_b/\bar{B^0}\approx0.5\!-\!1$ and $\Xi_b^0/\bar{B^0}$ ratio around 0.3 at very low transverse momentum, which are about $1.5$ larger then those of the corresponding charmed hadron ratios at the same collision energy. Furthermore, we discuss the relative ratios between charmed and bottomed hadrons, emphasizing how these observables can provide information about the distribution of charm and bottom quarks and, if experimentally observed, would further support the idea of quark-gluon plasma formation even in small collision systems.
\end{abstract}
\maketitle
\section{Introduction}
In relativistic heavy-ion collisions, experimental data collected at RHIC (Relativistic Heavy Ion collider) at BNL and LHC (Large Hadron Collinder) at CERN have shown the evidence for the creation of a new phase of matter known as quark-gluon plasma (QGP), as predicted by Lattice QCD calculations \cite{Borsanyi:2016ksw}. 
In recent years, heavy hadrons like $D$ mesons and $\Lambda_c$ baryons, containing heavy quarks or heavy antiquarks, serve as excellent probes for studying the QGP. 
Due to their large masses, the production of charm and bottom quarks can be computed through pQCD calculations, as implied by the QCD factorization theorem, that describes the production cross-sections of heavy hadrons at collider energies.
The transverse momentum spectra and the Nuclear modification factor ($R_{AA}$) of final heavy hadrons show a behaviour which is consistent with the presence of a QGP medium, and a non negligible interaction of heavy quark with the medium itself. 
The study of Heavy Flavor production can shed light on the heavy-flavor transport coefficients in the QGP, obtained studying the nuclear modification factor $R_{AA}(p_T)$ of D meson and the anisotropic flows $v_n(p_T)$, and also provides insights on the mechanisms of heavy flavor hadronization  in AA collisions. Over the past decades, several studies have explored the effects of interactions between heavy quarks and bulk particles \cite{Gossiaux:2008jv,He:2012df,Uphoff:2012gb,Cao:2015hia,Scardina:2017ipo,Das:2017dsh,Das:2016cwd,Das:2015ana,Plumari:2019hzp,Oliva:2020doe,Beraudo:2021ont,Sambataro:2022sns,Sambataro:2023tlv}. For a recent systematical comparison between different hadronization models in AA collisions see ref. \cite{Zhao:2023nrz}.
Heavy flavour hadronization, traditionally referred to as fragmentation in elementary collisions, can be an intrinsically soft process and thus relies on phenomenological modelling.
The assumption that charm (\textit{c}) and bottom (\textit{b}) fragmentation is universal across various collision systems has been a common assumption in literature. However, recent experimental findings in the study of charmed hadrons production have brought this universality hypothesis into question \cite{ALICE:2022exq,CMS:2023frs,LHCb:2018weo}. These experimental data indicate that charmed baryon over meson ratio is enhanced already in high multiplicity proton-proton (\textit{pp}) collisions at LHC energies \cite{ALICE:2020wfu,ALICE:2021psx}, w.r.t. elementary collisions like electron-positron ($e^- e^+$) and electron-proton ($e^- p$) collisions \cite{Lisovyi:2015uqa}.
In the mid-rapidity region of $pp$ and $pA$ collisions at LHC energies, measurements have revealed a surprising large production of $\Lambda_c$ baryons. At low $p_T$, this production results in $\Lambda_c/D^0 \sim 0.5\!-\!0.6$ which is nearly an order of magnitude larger than what is expected with standard fragmentation approach. Additionally, has been observed a notable $\Xi_c$ and $\Omega_c$ production, with a $\Xi_c/D^0 \sim 0.2\!-\!0.3$.
Moreover, recently, LHCb result in pp collisions at forward rapidity has reported a notably higher ratio of bottomed baryons to mesons, $\Lambda_b/\bar{B^0} \sim 0.4\!-\!0.5$, at intermediate transverse momentum compared to elementary $e^+ e^-$ collisions, that is about 0.2 \cite{LHCb:2019fns,LHCb:2023wbo}. Previous studies on bottom hadron production have primarily relied on a statistical hadronization model (SHM), assuming relative chemical equilibrium among various heavy hadron species \cite{He:2022tod}. These studies accounted for an expanded set of hadrons, including theoretically predicted excited states from quark model \cite{Ebert:2011kk} for bottom baryons, 
attributing the observed enhancement of  $\Lambda_b$ production, by  almost a factor 2, to the feed-down from these unobserved states. We have suggested that in the charm sector this increase may be attributed to the formation of a medium that favors a coalescence and fragmentation hadronization mechanism, even in small collision systems such as pp collisions at top LHC energies resulting in an enhancement of baryon production with respect to meson production \cite{Minissale:2020bif}.
Furthermore, these findings for small collision systems 
have been strengthen by the very recent studies with EPOS4-HQ \cite{Zhao:2023ucp} that 
suggest a strong correlation between heavy flavor production and the formation of Hot QCD matter.  
Recently another approach that use the in-medium hadronization mechanism in heavy ion collisions has been applied to study charm hadron production also in small systems, obtaining qualitatively good agreement with observed experimental heavy-flavor particle ratios \cite{Beraudo:2023nlq}.
This approach provides a complementary support to the idea that in \textit{pp} high multiplicity events the heavy hadron production can have the same origin as those observed in heavy-ion collisions.
In this Letter, we present predictions for bottom hadron production in $pp$ collisions at top LHC energies using a coalescence plus fragmentation approach.
Our predictions encompass the production of $B$, $B_s$, $\Lambda_b$, $\Xi_b$, and $\Omega_b$, which are likely to be measured in the near future. 
This approach holds the advantage of potentially providing a unified description of charm and bottom hadron production at low and intermediate $p_T$ in $pp$, $pA$, and $AA$ collisions above the TeV energy scale.
\section{Hadronization by coalescence and fragmentation}
\label{section:Hadronization_model}
The coalescence process, and its modelling, was originally proposed as a mechanism for hadronization applied to Heavy Ion collisions at RHIC energies, with the aim of explain the $p_T$ spectra, the baryon over meson ratio  and the splitting of elliptic flow between mesons and baryons \cite{Greco:2003xt, Fries:2003vb, Greco:2003mm, Fries:2003kq, Molnar:2003ff}. Successively, there has been extensions to incorporate finite width, considering off-shell effects \cite{Ravagli:2007xx, Ravagli:2008rt, Cassing:2009vt}. It has been applied at LHC energies to characterize the spectra of primary light hadrons such as $\pi, K, p, \phi, \Lambda$, and the baryon-to-meson ratios at both RHIC and LHC energies \cite{Minissale:2015zwa}. Furthermore, the coalescence model has been used to analyze heavy flavor hadron chemistry in $AA$ collisions \cite{Oh:2009zj, Plumari:2017ntm, Cho:2019lxb, He:2019vgs}, with recent extensions to include multi-charmed hadrons like $\Xi_{cc}$ and $\Omega_{ccc}$ \cite{Minissale:2023dct}.
In this section, we recall the basic elements of the coalescence model based on the Wigner formalism. 
In this approach the momentum spectrum of hadrons formed by coalescence of quarks can be written as:
\begin{eqnarray}
\label{eq-coal}
&&\frac{dN_{H}}{dyd^{2}P_{T}} = g_{H} \int \prod^{N_{q}}_{i=1} \frac{d^{3}p_{i}}{(2\pi)^{3}E_{i}} p_{i} \cdot d\sigma_{i}  \; f_{q_i}(x_{i}, p_{i})\nonumber \\ 
&\times& C_{H}(x_{1}...x_{N_{q}}, p_{1}...p_{N_{q}})\, \delta^{(2)} \left(P_{T}-\sum^{n}_{i=1} p_{T,i} \right)
\end{eqnarray}
The statistical factor $g_{H}$ represents the probability of forming a colorless hadron from quarks and antiquarks with spin 1/2. $d\sigma_{i}$ refers to an element of a space-like hypersurface, while $f_{q_i}$ denotes the quark (anti-quark) phase-space distribution functions for the i-th quark (anti-quark). 
The function $C_H$ is defined as $C_{H}(x_{1}...x_{N_{q}})=C^{N_q-1}f_{H}(x_{1}...x_{N_{q}}, p_{1}...p_{N_{q}})$ 
Where the Wigner function $f_{H}(x_{1}...x_{N_{q}}, p_{1}...p_{N_{q}})$ describes the spatial and momentum distribution of quarks in a hadron, and the factor $C^{N_q-1}$ is determined
to impose
that in the limit $p \to 0$ all the bottom quarks hadronize by coalescence in a bottomed hadron, so that the total coalescence probability gives $\lim_{p \to 0} P^{tot}_{coal}=1$.
In the formula above $N_{q}$ is the number of quarks composing the hadron. Eq.(\ref{eq-coal}) describes meson formation when $N_{q}=2$, and baryon formation when $N_{q}=3$. 
For $B$ mesons, the statistical factor $g_{M}=1/36$ represents the probability that two random quarks have the correct color, spin, and isospin matching the quantum numbers of the considered mesons. For the baryons studied in this analysis the statistical factor is $g_{B}=1/108$.\\
The Wigner density of the heavy hadron can be constructed from the hadron wave function. According to references \cite{Greco:2003vf,Oh:2009zj,Plumari:2017ntm}, the hadron wave function can be approximated by using the wave function of a three-dimensional harmonic oscillator state with the same root mean square radius. Therefore, the ground state of bottomed hadrons, we adopt a Gaussian shape in space and momentum for the Wigner distribution function 
\begin{eqnarray} 
  f_H(...)&=& \prod^{N_{q}-1}_{i=1} 8\exp{\Big(-\frac{r_{ri}^2}{\sigma_{ri}^2} - p_{ri}^2 \sigma_{ri}^2\Big)} 
\label{Eq:Wigner}
\end{eqnarray}
$N_{q}$ represent the number of constituent quarks where $N_{q}=2$ for mesons and $N_{q}=3$ for baryons.
The coordinate $\vec{x}_{ri}$ and $\vec{p}_{ri}$ are the relative coordinates in space and momentum in the CMS of the two particles. In the case of mesons, the relative coordinates ($x_{r1}$, $p_{r1}$) are given by 
\begin{eqnarray}
r_{r1}=|\vec{x}_1-\vec{x}_2|, ~ p_{r1}={|m_2\vec{p}_1-m_1\vec{p}_2| \over m_1+m_2},
\label{Eq:JACOBI1}
\end{eqnarray} 
while for baryons are defined as
\begin{eqnarray}
    r_{r1}=\frac{|\vec{x}_1-\vec{x}_2|}{\sqrt{2}}, ~ p_{r1}=\sqrt{2} \; {|m_2\vec{p}_1-m_1\vec{p}_2|\over m_1+m_2},
\end{eqnarray}  
and $r_{r2}$, $p_{r2}$  are given by
\begin{eqnarray}
r_{r2}&=& \sqrt{2 \over 3} \;  \left| {  {m_1\vec{x}_1+m_2\vec{x}_2\over m_1+m_2}}-\vec{x}_3 \right| , \nonumber\\
p_{r2}&=& \sqrt{3 \over 2} \;  {|m_3(\vec{p}_1+\vec{p}_2)-(m_1+m_2)\vec{p}_3|\over m_1+m_2+m_3} ,
\end{eqnarray}  
we notice that some authors employ 
$(E_{2} \vec{p}_{1}- E_{1} \vec{p}_{2})/(E_{1}+E_{2})$ that in the CMS is equivalent to Eq.\ref{Eq:JACOBI1}.
The covariant widths $\sigma_{ri}$ are related to the oscillator frequency $\omega$ by $\sigma_{ri}=1/\sqrt{\mu_i \omega}$ where $\mu_i$ are the associated reduced masses $\mu_1=(m_1 m_2)/(m_1+m_2)$ and $\mu_2= [ (m_1+ m_2)m_3]/[m_1+m_2+m_3]$.
The root mean square charge radius $\langle r^2\rangle_{ch}$ of the hadron is related to the covariant widths of the Wigner function $f_H$. 
For mesons, it is expressed as:
\begin{eqnarray} 
\langle r^2\rangle_{ch}&=& \frac{3}{2}  \frac{Q_1 m_2^2+Q_2 m_1^2}{(m_1+m_2)^2} \sigma_r^{2}
\end{eqnarray}
with $Q_i$ the charge of the i-th quark and the center-of-mass 
coordinate calculated as $X_{cm}=\sum_{i=1}^{2} m_i x_i/\sum_{i=1}^2 m_i$.
While in a similar way the oscillator frequency and the widths for baryons can be related to the root mean square charge radius of the baryons by
\begin{eqnarray} \label{Eq:rBaryon}
\langle r^2\rangle_{ch}&=& \frac{3}{2} \frac{m_2^2 Q_1+m_1^2 Q_2}{(m_1+m_2)^2} \sigma_{r 1}^2  \\ 
&+&\frac{3}{2} \frac{m_3^2 (Q_1+Q_2)+(m_1+m_2)^2 Q_3}{(m_1+m_2+m_3)^2} \sigma_{r 2}^2 \nonumber
\end{eqnarray}
In this approach the covariant widths are related by the oscillatory frequency $\omega$ through the reduced
masses by $\sigma_{p i}=\sigma_{r i}^{-1}=1/\sqrt{\mu_{i} \omega}$ therefore the model has only one free parameter that can be fixed by imposing the mean square charge radius of the hadron. 
The masses of light and heavy quarks used here are $m_{u,d}\!=\!300$ MeV, $m_{s}\!=\!380$ MeV, $m_{c}\!=\!1.5$ GeV and $m_{b}\!=\!4.8$ GeV.
The values of the mean square charge radius of mesons and baryons used in this work have been taken from quark model \cite{Hwang:2001th,Albertus:2003sx}. 
In Table \ref{table:param} we report the widths and the mean square charge radius considered.
\begin{table} [ht]
\begin{center}
  \begin{tabular}{l c c || l c c c}
    \hline
    \hline
    Meson &  $\langle r^2\rangle_{ch}$ & $\sigma_{p1}$ & Baryon & $\langle r^2\rangle_{ch}$ & $\sigma_{p1}$ & $\sigma_{p2}$ \\
    $B^{-}[b \bar{u}]$     & -0.378   & 0.302 & $\Lambda_b^0 [udb]$	   & 0.13   & 0.264  & 0.5 \\
    $\bar{B^{0}}[b \bar{d}]$     & 0.187   & 0.303 & $\Xi_b^0 [usb]$	  & 0.16   & 0.279  & 0.527 \\
    $\bar{B_{s}^{0}}[b\bar{s}]$   & 0.119   & 0.374  & $\Xi_b^- [dsb]$	  & -0.21   & 0.295  & 0.557 \\
    $B_{c}^{-}[b\bar{c}]$   & -0.043   & 0.74  &  $\Omega_b^- [ssb]$	  & -0.18   & 0.318  & 0.592 \\
\hline
\hline
\end{tabular}
\end{center}
\caption{Mean square charge radius $\langle r^2\rangle_{ch}$ in $fm^2$ and the widths parameters $\sigma_{pi}$ in $GeV$. The mean square charge radius are taken quark model \cite{Hwang:2001th,Albertus:2003sx}.}
\label{table:param}
\end{table}
Coalescence calculations have indicated that the probability of heavy quarks hadronizing through the coalescence mechanism in AA collisions decreases with the transverse momentum $p_T$ of the heavy quarks, thereby leaving the system to hadronize via fragmentation in the high $p_T$ region that must be included to describe the transition from low to high-momentum regime. 
The hadron momentum spectra from the heavy quark fragmentation is given by:
\begin{equation}
\frac{dN_{had}}{d^{2}p_T\,dy}=\sum \int dz \frac{dN_{fragm}}{d^{2}p_T\, dy} \frac{D_{had/c}(z,Q^{2})}{z^{2}} 
\label{Eq:frag}
\end{equation}
In our approach, we assume that heavy quarks that do not hadronize via coalescence with light quarks will hadronize by fragmentation with probability given by $P_{frag}(p_T)=1- P^{tot}_{coal}(p_T)$, that results in a parton distribution here indicated as $\frac{dN_{fragm}}{d^{2}p_T\, dy}$.
While $D_{\text{had/c}}(z,Q^{2})$ represents the fragmentation function, where $z=p_{\text{had}}/p_{c}$ denotes the momentum fraction transferred from heavy quarks to the final heavy hadron. The momentum scale for the fragmentation process is fixed by $Q^2=(p_{\text{had}}/2z)^2$. 
The final hadron spectra coming from fragmentation is evaluated by performing the convolution integral and integrating over all the momentum fraction $z$, between the momentum distribution of heavy quarks which do not coalesce and the Kartvelishvili fragmentation function \cite{Kartvelishvili:1977pi} given by $D(z) \propto z^ {\alpha}(1-z)$.
Here, $z = p_{had}/p_b$ represents the momentum fraction carried by the heavy hadron formed from the heavy quark fragmentation, and $\alpha$ is a parameter that can be fixed to reproduce the experimental HF mesons spectra in $pp$ collisions at higher transverse momentum $p_T$ measured at LHC \cite{CMS:2017uoy,CMS:2018eso}.
Following Ref\cite{He:2022tod}, we have employed a value of $\alpha=25$ for mesons and a value of $\alpha=8$ for baryons.
The fragmentation fraction that gives the probability that a heavy quark fragment in a specific heavy hadron, for B mesons is taken in accord with the world-average
production fractions of 40.2\% \cite{Workman:2022ynf} as used by CMS collaboration in Ref.\cite{CMS:2017uoy}. For all the others fragmentation fractions we have used values that are given in \cite{Oh:2009zj} which are consistent with the ones used by EPOS4HQ \cite{Zhao:2024ecc}. Notice that in these works the fraction for the mesons are in accord with the value aforementioned.
%
\section{Fireball and parton distribution}\label{section:Fireball}
The heavy quarks spectra in $pp$ collisions 
are described by hard processes and calculated by perturbative QCD (pQCD) at NNLO.
Therefore, in our calculation, the bottom quark spectrum have been taken in accordance with the bottom distribution in $p + p$ collisions with the Fixed Order + Next-to-Leading Log (FONLL), as given in Ref. \cite{Cacciari:2005rk,Cacciari:2012ny}.
The total bottom cross section used in this work is
$d\sigma_{b\bar{b}}/dy = 34.5 
 \mu b$, in accord with measurements of ALICE Collaboration \cite{ALICE:2021mgk}.
In small collision systems like \textit{pp} collisions at LHC energies, the effect on the heavy quark spectra  originating from in-medium HQ scattering is expected to be negligible  \cite{Song:2024hvv}. Furthermore, it is reasonable to assume that the modification of the spectrum due to the jet quenching mechanism could be negligible, especially considering that even in $pA$ measurements, an $R_{pA} \approx 1$ is observed. For the description of the QGP medium, we follow Ref. \cite{Minissale:2020bif} and assume a fireball dimension estimated in hydrodynamics and transport simulations \cite{Weller:2017tsr,Gallmeister:2018mcn}.
These models suggest a lifetime of the fireball at these collision energies of about $\tau \approx 2.5 , fm/c$ with a transverse radius of about $R_T\approx 2 fm/c$. Therefore, in our calculation the bulk of particles that we assume is a thermalized system of gluons and $u,d,s$ quarks and anti-quarks with longitudinal momentum distribution assumed to be boost-invariant in the range $y\in(-0.5,+0.5)$. We consider a radial flow with the following radial profile $\beta_T(r_T)=\beta_{max}\frac{r_T}{R_T}$.
\\
For the QGP medium we assume a mixture composed by low transverse momentum,  $p_T<2 \,\mbox{GeV} $ that are described by a thermal distribution
\begin{equation}
\label{quark-distr}
\frac{dN_{q,\bar{q}}}{d^{2}r_{T}\:d^{2}p_{T}} = \frac{g_{q,\bar{q}} \tau m_{T}}{(2\pi)^{3}} \exp \left(-\frac{\gamma_{T}(m_{T}-p_{T}\cdot \beta_{T})}{T} \right) 
\end{equation}
with transverse mass $m_T=\sqrt{p_T^2+m_{q,\bar{q}}^2}$. The spin-color degeneracy $g_{q}=g_{\bar{q}}=6$. Following  Ref.\cite{Biro:1994mp,Greco:2003mm,Minissale:2020bif} gluons are taken into account by converting them to quarks and anti-quark pairs according to the equilibrium flavour compositions.
For partons at high transverse momentum, $p_T> 2.5 \, \mbox{GeV}$ we consider a power law function as evaluated by pQCD calculations, see \cite{Plumari:2017ntm, Scardina:2010zz,Liu:2006sf}.
\\
\section{Bottomed hadron Transverse momentum spectra and ratios}\label{section:results}
In this section, we discuss the results for the transverse momentum spectra of $\bar{B^0}$, $B^-$, $B_s$ and $B_c$ mesons, and $\Lambda_b$, $\Xi_b^{0,-}$ and $\Omega_b$ baryons using the model described in previous sections, specifically for $pp$ collisions at top LHC energies.
\\
In this study we include ground state hadrons and the first excited resonances of $B$, $\Lambda_b$ and $\Xi_b^{0,-}$ as listed in
Table \ref{tab:bottom}, which include the states present in the Particle Data Group \cite{Zyla:2020zbs}.
Following the same approach used in \cite{Plumari:2017ntm} we consider the thermal statistical factor given by 
$[m_{H^*}/m_H]^{3/2} \times \exp {\left(-\Delta m /T\right)}$ with $\Delta m=m_{H^*}-m_H$, where
$m_{H^*}$ is the mass of the resonance. This statistical factor is given by the Boltzmann probability to populate an excited state of mass $m+\Delta m$, at a temperature $T$.
The inclusion of resonance decay gives an important contribution to the total particle yield in addition to the ground-state spectra.
\begin{table} [ht]
\begin{center}
\begin{tabular}{lcclcc}

\hline
Meson&Mass&I(J)& \vrule \vrule \, Baryon& Mass&I(J)\\
\hline 
$B^-[\bar{u}b]$		& 5280 & $\frac{1}{2} \,(0)$	& \vrule \vrule \, $\Lambda_b^0[udb]$	& 5620 & $0 \, (\frac{1}{2})$	\\
$\bar{B^0}[\bar{d}b]$ 	& 5280 & $\frac{1}{2} \,(0)$	& \vrule \vrule \, $\Xi_b^0[usb]$	& 5792 & $\frac{1}{2} \, (\frac{1}{2})$	\\
$\bar{B_{s}^{0}}[\bar{s}b]$	& 5366 & $0 \,(0)$		& \vrule \vrule \, $\Xi_b^-[dsb]$	& 5797 & $\frac{1}{2} \, (\frac{1}{2})$	\\
$B_{c}^{-}[\bar{c}b]$	& 6275 & $0 \,(0)$		& \vrule \vrule \, $\Omega_b^-[ssb]$	& 6045 & $0 \, (\frac{1}{2})$	\\
\hline
\hline
Meson Res.& Mass& I(J)& Decay& B.R.\\
\hline
$B^* $	& 5325 & $\frac{1}{2} \, (1)$	& $B\gamma$\\
$B_1(5721) $	& 5726 & $\frac{1}{2} \, (1)$		& $B^*\gamma$ & $100\%$\\
$B_2(5747) $	& 5737 & $\frac{1}{2} \, (2)$		& $B^*\gamma$ & $100\%$\\
$B_s^* $	& 5415 & $0 \,(1)$		& $B_s\gamma$ & \\
$B_{s1}^0(5830) $	& 5829 & $0 \,(1)$		& $B^{*+}K^-$ & \\
$B_{s2}^0(5840) $	& 5840 & $0 \,(2)$		& $B^+K^-$ & \\
\hline
\multicolumn{2}{l}{Baryon  Res.}\\
\hline
$\Lambda_b^0(5912)$	& 5912 & $0 \, (\frac{1}{2})$	& $\Lambda_b^0 \pi^+\pi^-$ & $100\%$ \\
$\Lambda_b^0(5920)$	& 5920 & $0 \, (\frac{3}{2})$	& $\Lambda_b^0 \pi^+\pi^-$ & $100\%$ \\
$\Lambda_b^0(6070)$	& 6072 & $0 \, (\frac{1}{2})$	& $\Lambda_b^0 \pi^+\pi^-$ & $100\%$ \\
$\Lambda_b^0(6146)$	& 6146 & $0 \, (\frac{3}{2})$	& $\Lambda_b^0 \pi^+\pi^-$ & $100\%$ \\
$\Lambda_b^0(6152)$	& 6152 & $0 \, (\frac{5}{2})$	& $\Lambda_b^0 \pi^+\pi^-$ & $100\%$ \\
$\Sigma_b$	& 5811 & $1 \, (\frac{1}{2})$	&$\Lambda_b^0 \pi$ & $100\%$\\
$\Sigma_b^*$	& 5830 & $1 \, (\frac{3}{2})$	&$\Lambda_b^0 \pi$ & $100\%$\\
$\Xi_b^{'\;-}(5935)$	& 5935 & $(\frac{1}{2})$	&$\Xi_b^{0} \pi^-$ & $100\%$\\
$\Xi_b^{0}(5945)$	& 5952 & $(\frac{3}{2})$	&$\Xi_b^{-} \pi^+$ & $100\%$\\
$\Xi_b^{-}(6100)$	& 5952 & $(\frac{3}{2})$	&$\Xi_b^{-} \pi^-\pi^+$ & $100\%$\\
\hline
\end{tabular}
\end{center}
\caption{
  Heavy hadrons with b quarks content considered in this work. We report the ground states and the first exited states
  including their decay modes with their corresponding branching ratios as given in Particle
  Data Group \cite{Zyla:2020zbs,Agashe:2014kda}.
\label{tab:bottom}}
\end{table}
\\
\begin{figure}[t]
\centering
\includegraphics[width=\columnwidth, angle=0,clip]{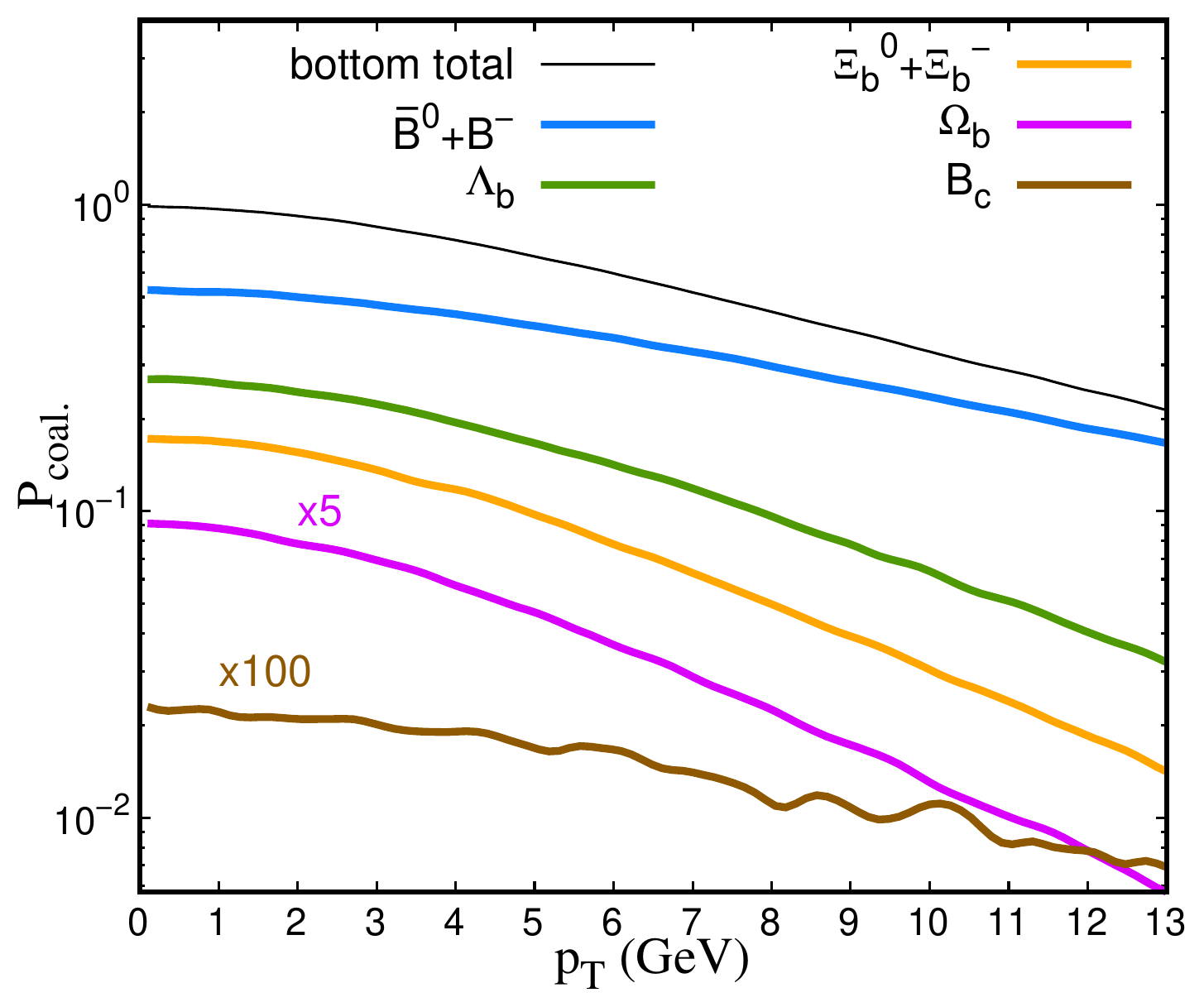}
\caption{
(Color online) Quark coalescence probabilities as a function of the 
transverse momentum of the bottom quark for $pp$ collisions at LHC energies.
The blue, green, maroon, red, orange and yellow lines are the coalescence probabilities to produce $B^{0}$, $B_{s}$, $\Lambda_b$, $\Xi_b^{0,-}$, $\Omega_b$ and $B_c$ respectively. Black dashed line is the total probabilities that a b quarks hadronize by coalescence in heavy meson or baryon.
}
\label{Fig:Pcoal}
\end{figure}
In Fig.\ref{Fig:Pcoal}, we show the coalescence probabilities $P_{coal}$ for bottom quarks to hadronize into specific hadrons ($\bar{B^{0}}+B^-$, $B_{s}$, $\Lambda_b$, $\Xi_b^{0}+\Xi_b^{-}$, $\Omega_b$, and $B_c$) as a function of the bottom transverse momentum. As illustrated, $P_{coal}$ decreases with increasing $p_T$, reflecting the trend observed for charm quarks in \cite{Minissale:2020bif} but with a quite wider extension.  
Consequently, at high $p_T$, fragmentation emerges as the dominant mechanism for bottom hadronization, with bottom quarks transitioning into specific final bottom  hadron channels based on different fragmentation fractions, as outlined in Ref. \cite{Oh:2009zj}.
\\
The comparison of the coalescence probabilities for a charm quark \cite{Minissale:2020bif} and a bottom quark reveals a similar dependence on the heavy quark's transverse momentum. However, the slope for bottom quarks is smaller than that for charm quarks, indicating that coalescence is expected to exert a greater influence on bottom hadron production for a given transverse momentum.
A quantitative comparison between the two heavy quark coalescence probability shows that at $2 \; \text{GeV}$ a charm quark has about 50\% probability to hadronize via coalescence process, while a bottom quarks with the same momentum has a probability of about 90\%.
At $p_T \sim 7 \text{GeV}$ a bottom quark has a probability of about 50\%, but for a charm quark at that momentum the probability is of about one order of magnitude smaller.
In particular it can be seen that while $D^0$ meson is dominated by fragmentation at all $p_T$, 
for $\bar{B^0}$ one has a dominance of the coalescence contribution up to $p_T \approx 6 \text{GeV}$.
\\
Furthermore, similar to the findings for charmed hadrons in \cite{Minissale:2020bif}, the comparison of different coalescence probabilities in Fig. \ref{Fig:Pcoal} reveals that at lower momenta, the coalescence probability for $\Lambda_b$ and $\Xi_b^{0,-}$ is similar to the $\bar{B^0}$ probability. This behaviour of the coalescence mechanism suggests an enhancement of the $\Lambda_b/\bar{B^0}$ and $\Xi_b^{0,-}/\bar{B^0}$ ratios with respect to elementary collision.
\begin{figure}[h!]
\centering
\includegraphics[width=\columnwidth, angle=0,clip]{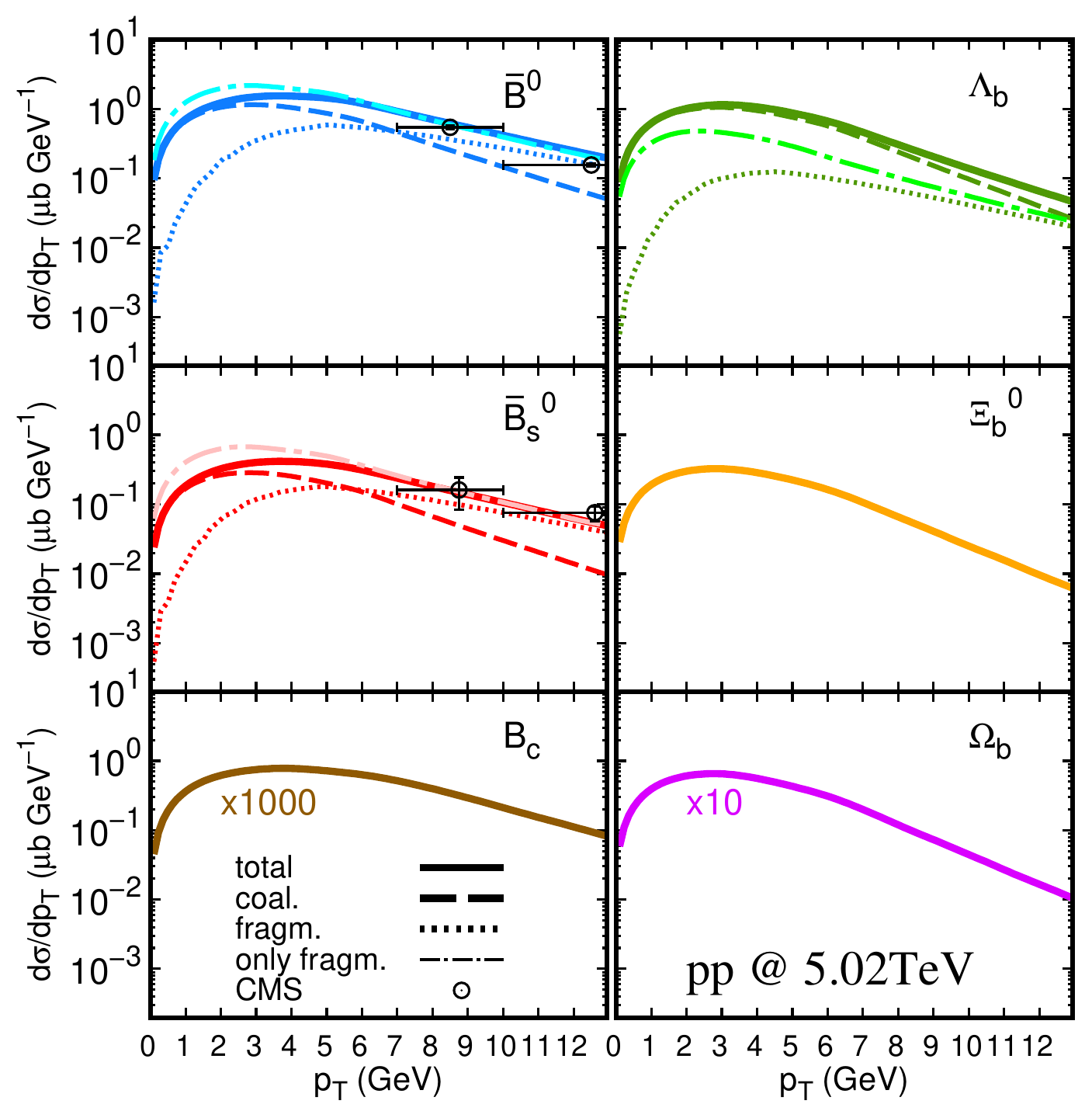}
\caption{
\label{Fig:spectra}
(Color online) Transverse momentum differential cross section for $\bar{B^0}$, $B_s$ mesons and $\Lambda_b$,$\Xi_b^{0,-}$ and $\Omega_b$
baryons at mid-rapidity for $pp$ collisions at $\sqrt{s}=5 \, \mbox{TeV}$.
Dashed and dotted lines refer to the spectra contribution from coalescence and
fragmentation, respectively; Solid line refers to the sum of
fragmentation and coalescence processes. Dot-dashed lines corresponds to the picture where fragmentation is the only hadronization mechanism. Data are from CMS Coll. \cite{CMS:2017uoy}
}
\end{figure}
In Fig. \ref{Fig:spectra} we show the $p_T$ spectra of $\bar{B^0}$ (left-top panel), $B_s$ (left-mid panel), $B_c$ (left-bottom panel), $\Lambda_b$ (right-top panel), $\Xi_b^{0}$ (right-mid panel) and $\Omega_b$ (right-bottom panel) for $pp$ collisions at mid-rapidity. The dashed lines
and the dotted lines refer, respectively, to the hadron spectra obtained by the contribution from coalescence and fragmentation. For $B_c$, $\Xi_b^{0}$ and $\Omega_b$ we have considered only the coalescence contribution.
Differently from what was obtained for $D^0$ mesons we observe a contribution from coalescence that is dominant for the production of $\bar{B^0}$ (and $B^-$, that have no significant difference with $\bar{B^0}$) in the low and intermediate transverse momentum range $p_T\le 6\!-\!7 \, GeV$ while fragmentation gives only a few percent of contribution to the total meson spectrum. \\
As shown in Fig.\ref{Fig:spectra} (right panels) the coalescence mechanism is the dominant mechanism for the baryon production, similarly to what obtained in pp and AA collisions. In particular for the $\Lambda_{b}$ production, in the whole momenta range considered, the coalescence production is larger than the fragmentation contribution. The $\Lambda_{b}$ is the only baryon that we have considered for the bottom fragmentation, with a fragmentation fraction of about $8\%$ of bottom quarks.
As well known, the baryon production enhancement is another peculiar feature of the recombination mechanism (see \cite{Minissale:2015zwa,Plumari:2017ntm,Minissale:2020bif}) that favours the process of massive hadron formation when a QGP medium is present (that we assume to be formed also in collision system like $pp$) w.r.t. large mass particle production via fragmentation. 
Furthermore, in Fig. \ref{Fig:spectra} with dot-dashed lines we show also the results for bottom mesons and baryon production assuming a scenario where fragmentation is the only hadronization mechanism. As shown, one is able to describe the momentum spectrum of $\bar{B^0}$ at high $p_T$ and an enhancement of $\bar{B^0}$ at low transverse momentum as compared with the solid line. Hence, looking at the sole $\bar{B^0}$ spectra, the differences between the two scenarios, in light of the current experimental data, cannot be sufficiently distinguished.
On the other hand, the difference in the case of baryons are significant, in fact in an approach where the fragmentation is the only mechanism of hadronization the $\Lambda_b$ production is lower of about a factor 2 in the region around $p_T\sim 2\!-\!10 \;\text{GeV}$ w.r.t. the case when also coalescence is taken into account. We will see in the following section the consequences on the baryon over meson ratios coming from the choice of assuming only fragmentation or the contribution of both coalescence and fragmentation.
We have extended the study on the particle production considering also single-bottomed hadrons with strange and charm content. The first group of particle includes the $B_s$ meson and the $\Xi_b^{0,-} \; \text{and} \; \Omega_b$ baryons; for the second one we have the $B_c$ meson.
For the $B_s$ spectrum we obtain about a factor about 1/4 smaller than the $\bar{B}^0$, in this spectrum the contribution of both mechanism are comparable in the $p_T$ region around $5 \text{GeV}$.
The $\Xi_b^{0}$ (and similarly for $\Xi_b^{-}$) $p_T$ differential cross section is about a factor 3 smaller than the $\Lambda_b$ cross section; in this case we have considered only the coalescence contribution, because the fragmentation fraction is expected to be negligible (in \cite{He:2022tod} the fragmentation fraction is estimated to be near 3\%).
The $\Omega_b$ cross section is about one order of magnitude smaller than the $\Lambda_b$ and $\Xi_b^{0,-}$ one, and as done for the $\Xi_b^{0,-}$ we have considered only the contribution from coalescence. For all these baryon species we don't see between them a significant difference in the slope as a function of momentum.
The $B_c$ production results of about two order of magnitude smaller than the mesons with light quarks content.
\begin{figure}[t]
\centering
\includegraphics[width=\columnwidth, angle=0,clip]{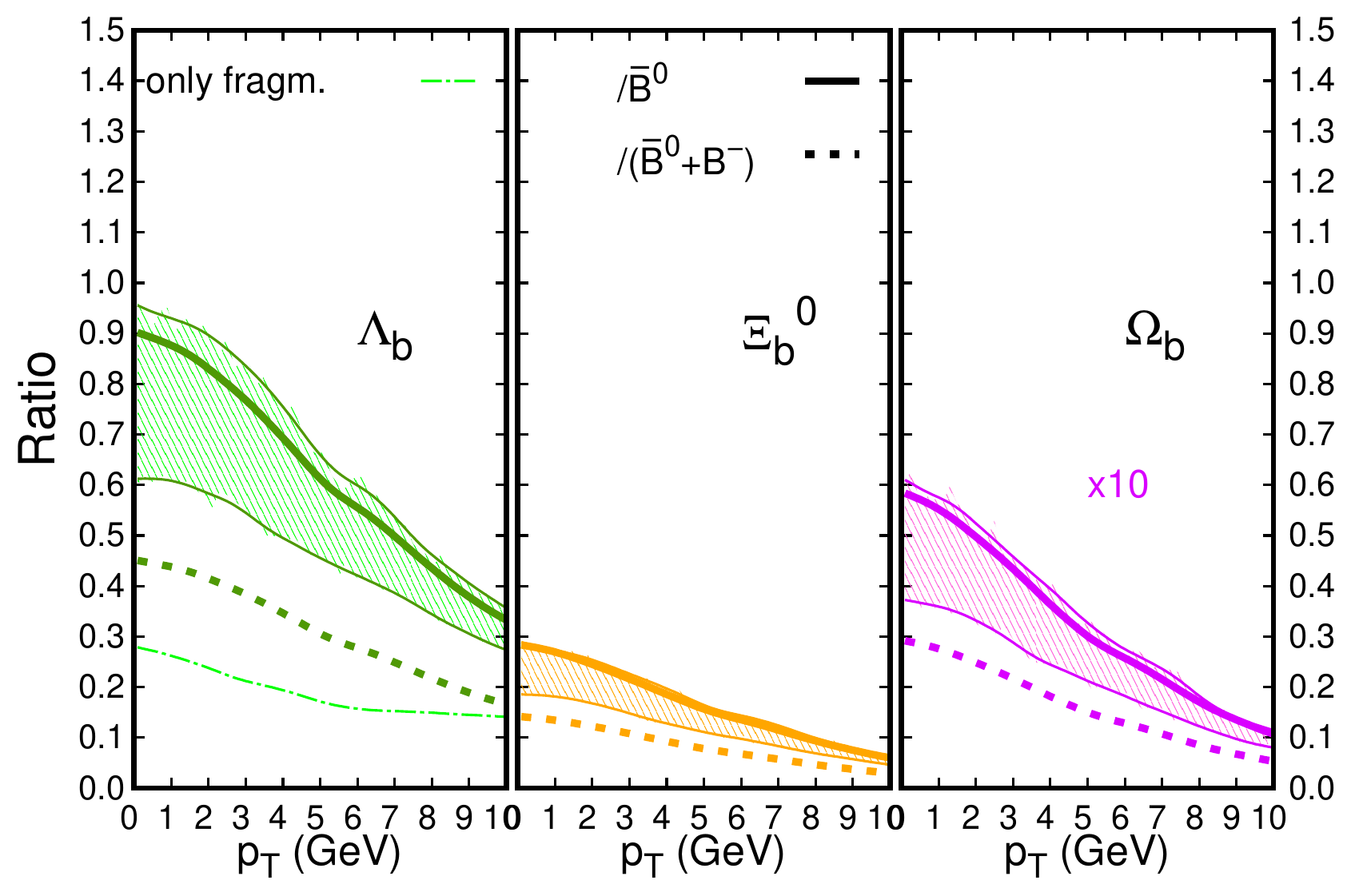}
\caption{
\label{Fig:ratioLbB}
(Color online) 
Ratios of $\Lambda_{b}$ (green), $\Xi_{b}$ (orange), $\Omega_{b}$ (purple) with $\bar{B^0}$ (solid line) and $\bar{B^0}+B^-$ (dashed line) as a function of $p_T$ and at mid-rapidity for 
$pp$ collisions at $\sqrt{s}=5 \, \mbox{TeV}$. Dot-dashed line correspond to the case where the fragmentation is the only process. The bands correspond to the sensitivity of the ratio with a variation of the Wigner widths. 
}
\end{figure}
\\
In \textit{pp} collisions the heavy hadron production has been usually considered as the one present in elementary collisions like $e^+e^- \; \text{and} \; e^-p$, described by the fragmentation and their fractions that determine the baryon over meson ratio. But the ratios in \textit{pp} collisions turn out to be well larger than the one expected from the solely fragmentation as pointed out by several experimental evidence \cite{ALICE:2021psx,ALICE:2020wfu}.
In the following we discuss the spectra ratio between bottom hadrons as a function of transverse momentum.
In Fig. \ref{Fig:ratioLbB}, we present results for the  baryon over meson ratio of $\Lambda_{b}$, $\Xi_b^0$ and $\Omega_b$ baryons in the case of $\bar{B^0}$ meson and with the sum $\bar{B^0}+B^-$,
for $pp$ collisions at $\sqrt{s}=5.02 ;\text{TeV}$. The hybrid coalescence plus fragmentation approach predict for bottomed baryon a larger ratio than the one observed in the charm sector. 
The uncertainties coming from the mean square charge radius in the Quark Model \cite{Albertus:2003sx,Hwang:2001th} corresponds to a variation of about 10\% for the Wigner functions widths $\sigma_r$. 
We have considered a uncertainty of a $\pm10\%$ widths variation to get an indicative value of the impact of this model parameter. 
We have calculated the scenario where we increase all the baryons radii and decrease all the mesons radii (upper limit), and viceversa (lower limit). So the upper and lower limit are the extreme possible scenario on the baryon over meson ratio. 
The coalescence mechanism significantly enhances $\Lambda_{b}^{0}$ with $\Lambda_{b}/B^{0} \approx 0.9$ at low $p_T$ which is about a factor 2 larger than the one of $\Lambda_c/D^0$. 
The ratio decreases with increasing $p_T$, which is a behaviour observed also for charmed baryon in \textit{pp} collision both experimentally and theoretical.
For $\Xi_b^0/\bar{B^0}$ ratio (middle panel) the coalescence mechanism predict a ratio $\sim 0.3$ at very low momenta, which is smaller of about a factor three with respect to $\Lambda_b/\bar{B^0}$ but comparable with the experimental data and the prediction provided in Ref. \cite{Minissale:2020bif} for the corresponding charmed ratio $\Xi_c/D^0$ at the same collision energies. The $\Omega_b/\bar{B^0}$ ratio is $\sim 0.06$ at low momenta (in Fig.\ref{Fig:ratioLbB} it is multiplied by a factor 10) and has a similar trend in $p_T$ as the previous two baryon over meson ratios. Also for $\Omega_b$ we obtain a $\Omega_b/\bar{B^0}$ ratio that is comparable with the prediction of $\Omega_c/D^0$.
\begin{figure}
\centering
\includegraphics[width=\columnwidth, angle=0,clip]{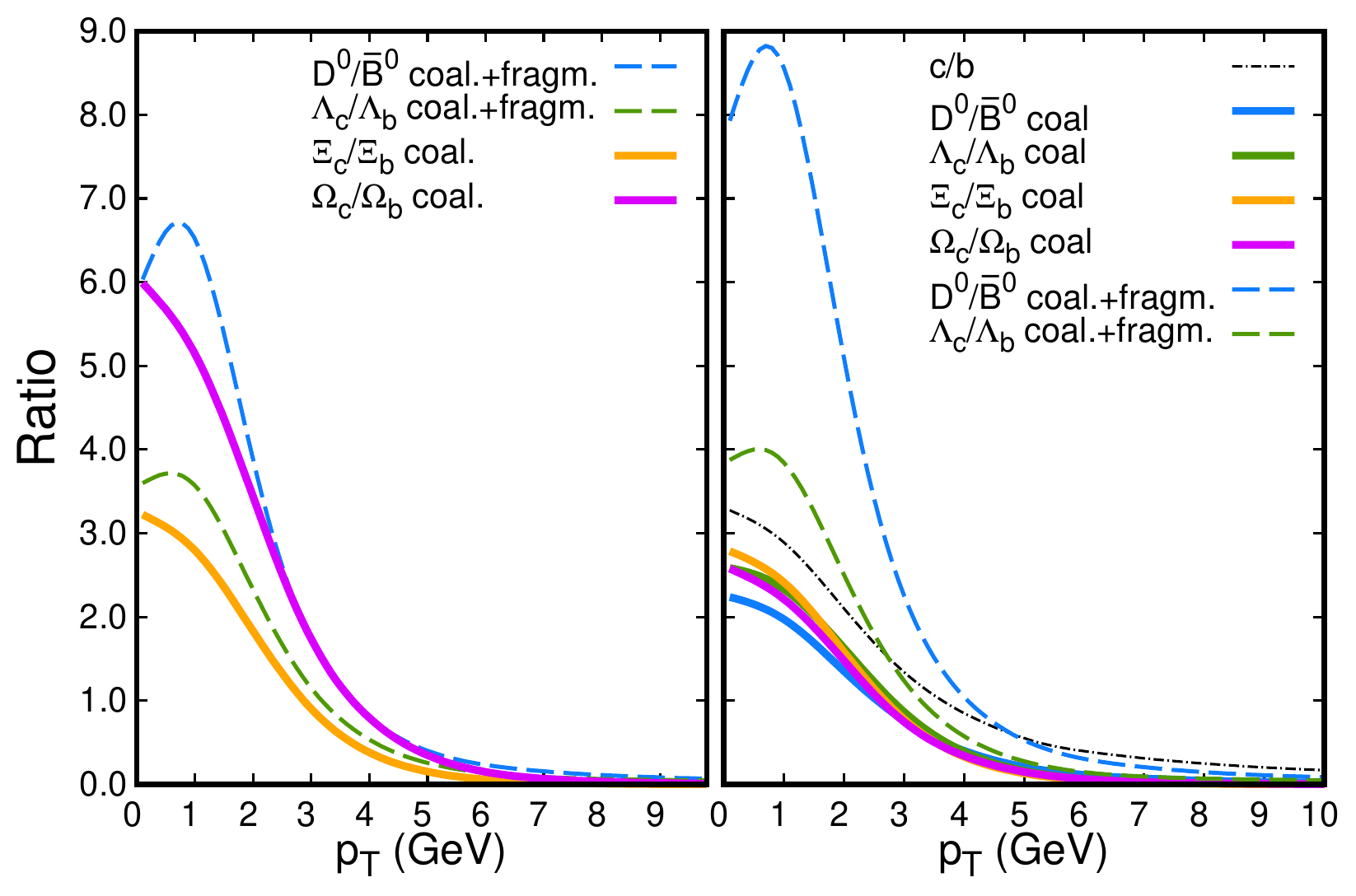}
\caption{
\label{Fig:ratio_cb}
(Color online) [left] charmed/bottomed hadrons respectively for $D/B$, $\Lambda_c/\Lambda_b$, $\Xi_c/\Xi_b$ and $\Omega_c/\Omega_b$ as a function of $p_T$ and at mid-rapidity for 
$pp$ collisions at $\sqrt{s}=5 \, \mbox{TeV}$.
[right] charmed/bottomed hadrons scaled with degeneracy and resonances factor, and \textit{c/b} quark ratio (black dot-dashed line).
}
\end{figure}
Finally, we discuss further information that one can gain about the microscopic details of hadronization by comparing charmed and bottomed hadrons spectrum.
In Fig. \ref{Fig:ratio_cb}, we show in the left panel the ratios of charmed hadrons over the corresponding bottomed hadrons ($H_c/H_b$), specifically $D/B$, $\Lambda_c/\Lambda_b$; alongside the corresponding hadrons with strange quarks content $\Xi_c/\Xi_b$ and $\Omega_c/\Omega_b$. In the right panel we show the aforementioned ratios scaled with the hadron and resonances degeneracy and suppression factors. In our model this correspond to consider the direct contribution of the ground state particle.  
Assuming a scenario where the coalescence is the only possible hadronization mechanism, a rough estimation employing Eq.\ref{eq-coal} should predict that the ratio $H_c/H_b$ should be proportional to the ratio of charm over bottom quark spectrum. From this consideration one would expect an approximate scaling of these ratios, except for effects arising from the convolution of distribution functions with the respective Wigner functions and from  resonance decay. The adoption of an hybrid mechanism by coalescence plus fragmentation tends to destroy such approximate scaling.
As discussed for the $\Xi_c$, $\Xi_b$ and $\Omega_c$, $\Omega_b$ baryons, we expect that the contribution from fragmentation to be negligible at least at low $p_T$ ($\leq \!6\!-\!8 \text{ GeV}$). This is in contrasts with what is expected for the $D$, $B$ mesons and $\Lambda_c$ and $\Lambda_b$ baryons. This suggest that one should expect an approximate scaling for the $\Xi$ and $\Omega$, reflecting the underlying microscopic structure, and a different ratio between $D/B$ and $\Lambda_c/\Lambda_b$ due to the contribution of fragmentation in their production mechanism.
In the left panel of Fig.\ref{Fig:ratio_cb} we show these ratios as a function of transverse momentum. The dashed lines refer to the ratios of $D/B$ and $\Lambda_c/\Lambda_b$ with the contribution of both coalescence and fragmentation. 
According to the the discussion above, the scaling of \textit{c/b} baryons with \textit{c/b} quarks from coalescence should be visible for those hadrons where fragmentation is negligible. In fact, in the right panel of Fig.\ref{Fig:ratio_cb} we have shown with solid lines the ratios $H_c/H_b$ for the case with only coalescence and with dashed lines the ratios where we have both coalescence and fragmentation. The black dotted line refer to the ratio of charm over bottom quark spectrum. We can see that in the case of only coalescence, all the hadrons considered in this work show a scaling behaviour with a function which is proportional to the ratio of charm/bottom quarks (dotted line). On the other hand, when we include both coalescence and fragmentation bewtween the $D/B$ and $\Lambda_c/\Lambda_b$ (blue and green dashed lines) we get a difference of about a factor 2.
These results show that these ratios can be an observable that could provide a way to discriminate between the two different hadronization processes, and at the same time can reflect on hadron level the behaviour of the distribution function at charm and bottom level.
In fact we notice that $\Lambda_c/\Lambda_b$ is predicted to be quite closer to $c/b$ w.r.t. $D/B$ because the former in our approach are quite dominated by coalescence. 

\section{Conclusions}
In this letter we have studied the production of
bottom hadrons in \textit{pp} collisions, calculating the baryon and meson spectra and their ratios by employing an hybrid hadronization mechanism via coalescence and fragmentation. 
Using our hybrid approach we have found a good description of the $\bar{B^0}$ and $B_s$ mesons with the experimental data from CMS collaboration at high $p_T$ \cite{CMS:2017uoy}. 
Furthermore, in this work we provide predictions for $B_c$ meson and $\Lambda_b$, $\Xi_b$ and $\Omega_b$ baryons. 
In a similar way as observed in the charm sector \cite{Minissale:2020bif}, the baryon production results enhanced in \textit{pp} collisions w.r.t. the simple production from fragmentation. 
An enhancement of the baryon over meson ratio for both bottom and charm hadrons support the idea that in small collision systems a QGP phase is formed. 
The results shown in this paper suggest that the presence of a hot and dense QCD matter in small collision systems permits the recombination of bottom quarks that significantly modify and enhance the bottom baryon production, and showing, from low to intermediate $p_T$, a dominant coalescence contribution  for $B$ mesons, differently for what found for $D$ mesons \cite{Minissale:2020bif}. 
From the comparison with the charmed hadrons, in the same collision system, we observe that the study of the relative hadron ratios $H_c/H_b$  can provide information about the distribution of charm and bottom quarks in \textit{pp} collisions providing at the same time insight on the interplay between coalescence and fragmentation. In the $p_T$ region where $\Xi_c$ and $\Xi_b$ or $\Omega_c$ and $\Omega_b$ are dominated by coalescence their ratio should reflect the evolution with $p_T$ of the $c/b$ ratio.

\section{Acknowledgments}
V.G. and S.P. acknowledges the funding from UNICT under ‘Linea di intervento 2’ (HQCDyn Grant). The authors also acknowledge the support from the European Union's Horizon 2020 research and innovation program Strong 2020 under grant agreement No 824093 and PRIN2022 (Project code 2022SM5YAS) within Next Generation EU fundings.

\end{document}